# Modeling and Evaluating Performance of Routing Operations in Reactive Routing Protocols

D. Mahmood[1], N. Javaid[1], Z. A. Khan[2], U. Qasim[3] , A. Khan[1], S. Qurashi[4], A. Memon[5]

[1]COMSATS Institute of Information Technology, Islamabad, Pakistan.
[2]Internetworking Program, Faculty of Eng., Dalhousie University, Halifax, Canada.
[3]University of Alberta, Alberta, Canada.
[4]IT Section, Hamdard University, Islamabad, Pakistan.
[5]IT Section, Associated Press of Pakistan, Islamabad, Pakistan.

## I. ABSTRACT

**Reactive routing protocols are gaining popularity due to their event driven nature day by day. In this vary paper, reactive routing is studied precisely. Route request, route reply and route maintenance phases are modeled with respect to control overhead. Control overhead varies with respect to change in various parameters. Our model calculates these variations as well. Besides modeling, we chose three most favored reactive routing protocols as Ad-Hoc on Demand Distance Vector (*AODV*), Dynamic Source Routing (*DSR*) and Dynamic *MANET* on Demand (*DYMO*) for our experiments. We simulated these protocols using *ns-2* for a detailed comparison and performance analysis with respect to mobility and scalability issues keeping metrics of throughput, route delay and control over head. Their performances and comparisons are extensively presented in last part of our work.**

## I. INTRODUCTION

Communication is one of the major needs of mankind. To receive or send any information, we need some communication network. Gradually, reaching to excellence, concept of Wireless Multihop Networks (*WMhNs*) gives enough liberty of freedom in this aspect. Considering such networks, each nod besides doing its prescribed job also act as a routing device along with being a transceiver. Information coming from one node is passed uninterrupted to next node till it reaches its destination. More over, these networks can extend up to thousands of nodes as in wireless sensor networks or need very efficient routing as in body area networks where packet drop ratio must tends to zero, or these networks may have high mobility as defined in vehicular ad-hoc networks. These all constraints, major of which are scalability and mobility, are still open research issues and lots of work is in progress [1].

To achieve such goals, we need some efficient protocols for network layer in *OSI* model. Major concern of network layer protocol is to establish, look after and give synchronization amongst all possible routes of network. Hence it can easily be stated as network performance is dependant on efficiency of routing protocol ([2],[3]). Extensive work has been done in this aspect (e.g. [4],[5],[6]) and today there are three major categories of network layer routing protocols for wireless multihop networks naming, reactive routing protocols, proactive routing protocols and hybrid routing protocols. In this paper, we are concerned only with reactive routing of wireless multihop networks. This category of protocols as name indicates is based on event occurrence. As, a node needs to transmit some data to a desired destination, reactive protocol, at that instance starts searching its route. Nodes that are in way to destination node act as relays or routers. Three prominent reactive protocols i.e. DYnamic MANET On-demand ($DYMO$) [7]), Ad-hoc On-demand Distance Vector ($AODV$) ([8],[16]) and Dynamic Source Routing ($DSR$) [9],[17]) are under consideration. These protocols are studied for producing mathematical framework of control over head. This framework is extended for a network having variations in different network parameters. Finally extensive simulations under mobile and scalable environments are conducted to present their comparisons and performance analysis with respect to different network metrics.

## II. RELATED WORK

Reactive Routing is not a very new concept and so, a lot of research is conducted on this aspect (e.g. [10], [11], [12]). Modeling Routing overhead is another step ahead for betterment of reactive routing ([13],[15],[17]). Besides comparing different routing protocols to give appropriate protocol for appropriate environment is also helpful [18]. Considering existing work done on this subject, ([19],[21]) provide analytical framework for calculating routing overhead of reactive protocols. They quantify route discovery process, i.e., overhead due to route $REQuest$ packets and route $REPly$ packets of any network underlying a reactive routing protocol. However, link monitoring overhead is not considered in their work. Authors of [22] give a combined framework of reactive and proactive routing protocols. The proposed models address scalability issues of a network. An analytical model which presents the effect of traffic on routing overhead was proposed by [23], whereas, [24] presents a survey of routing overhead on both reactive and proactive protocols and discuss cost of energy as routing metric.

I.D Aron *et.al* [25] present link repairing modeling, both in local repairing and source to destination repairing of two routing protocols, which were $DSR$ and $WRP$. They compare these two routing protocols, though aggregate routing overhead is not considered in [25]. In [26], authors present brief understanding of scalability issues of network; however, impact of topology change is not sufficiently addressed. [36] Very effectively presented a programming model for reactive routing with respect to mobile Ad-Hoc Networks. In parallel to [36], [37] produced excellent model representing network connectivity. His work is applicable for both $MANETs$ and $VANETs$. N. Javaid et al. [39] ensured energy and delay of reactive routing for wireless multi-hop network. Authors of ([39],[14]) give detailed analysis on performance metrics of $MANETs$ and $VANETs$ routing Protocols. Kumar. S et al. [43] addresses path stability and link duration in reactive as well as proactive routing protocols for $MANETs$. They chose, $DYMO$ as reactive routing protocol where as $OLSR$ and $DSDV$ were studied as proactive routing protocols. [34] produces excellent framework covering control overhead for proactive routing protocols. In this work, authors have presented mathematical model for generalized control overhead and discussed three most wanted proactive routing protocols in brief. Considering Vehicular ad-hoc networks, [36] modified a reactive routing protocol ($DSR$) and two proactive routing protocols ($FSR$ and $OLSR$). In general sense, they modified mobility and scalability aspects in link state routing for $VANETs$. Expanding Ring Search and Binary Back off Algorithms are prominent algorithms to reduce routing overhead. In [45] authors modeled and modified $ERS$ algorithm for $AODV$ and $DSR$. [43] Very clearly narrated mobility issues concerning wireless multi hop networks. We in [46] give detailed overview, operations and mathematical modeling of routing overhead with respect to $AODV$, $DSR$ and $DYMO$.

We enhanced our presented framework of reactive routing in [40] discussing over all overhead by route discovery and route maintenance processes. In this paper, we modified our existing work by giving a detailed comparison analysis on functionality, operability and performance of chosen three routing protocols of reactive in nature, i.e. *AODV, DSR, DYMO*. This discussion is presented in graphical, textual as well as tabular form to get better understanding of these routing protocols. In this work we have discussed mobility as well as scalability aspect of a network in detail.

## III. REACTIVE ROUTING

Unlike proactive routing protocols, where all the routes are formulated whenever the network initializes (e.g. [39],[40],[41]), in reactive approach, routes are queried only when needed by a node. A route request is flooded in the entire network and when a route is established, data is to be sent. Route discovery and route maintenance are the two major aspects of routing overheads of a reactive routing protocol [38].

### A. Route Discovery

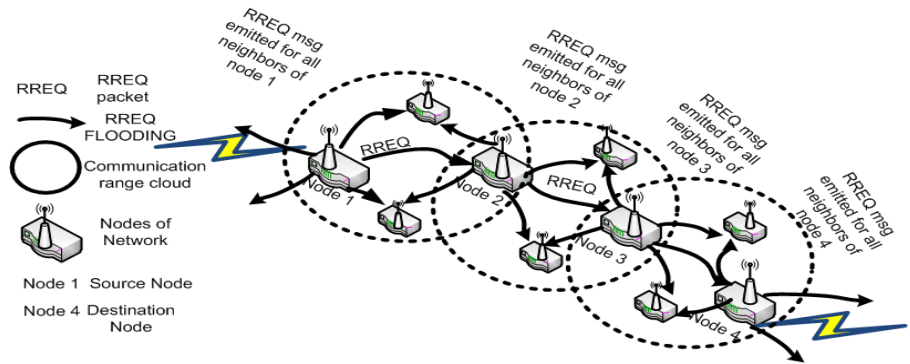

Figure 1: Route Request Process (RREQ)

The process of $RREQ$ packet propagation is shown in fig. 1. When a node requires a route, $RREQ$ packet is propagated until it reaches destination node. This propagation can be uncontrolled or flooded where as, in $AODV$ and $DSR$, expanding ring search algorithm limits the control overhead that can be generated via uncontrolled flooding of route request message.// An $RREQ$ packet comprises of many fields, most prominent ones are Source identifier field to identify the route requesting node, Destination identifier field to identity the destination node and $TTL$ field to limit the flooding or other purposes that can be defined according to need of the protocol. This uniquely identified $RREQ$ is flooded amongst all the nodes of network until it reaches destination node via different paths/ routes. Via $RREQ$ packet which has reached destination node, destination node will keep the route back to source and send an $RREP$ packet to all those routes from which $RREQ$ has reached it. On every node, where packet reaches, hop count / $TTL$ is incremented / decremented and route table entries are updated [16].

Normally it is expected that there is a bidirectional communications between originator and destined node i.e. not only source should know the route to its destination but destination should also know the route to source. For this purpose, as shown in fig. 2, a $RREP$ packet is generated. Only that node can generate $RREP$ message that itself is destination, or that has a valid route to the destination. If the route is discovered and $RREQ$ packet has reached at a node which has fresh route to destination or destined node itself than, an $RREP$ packet is generated. Via reverse path, $RREP$ packet is routed back while the nodes on this route establish the forward path entries in their routing tables. These entries than finally provide an active forward route to the destination from source. To avoid stale routes there is a route timer associated with each route/ path entry. Whenever the timer expires, the route is deleted [16].

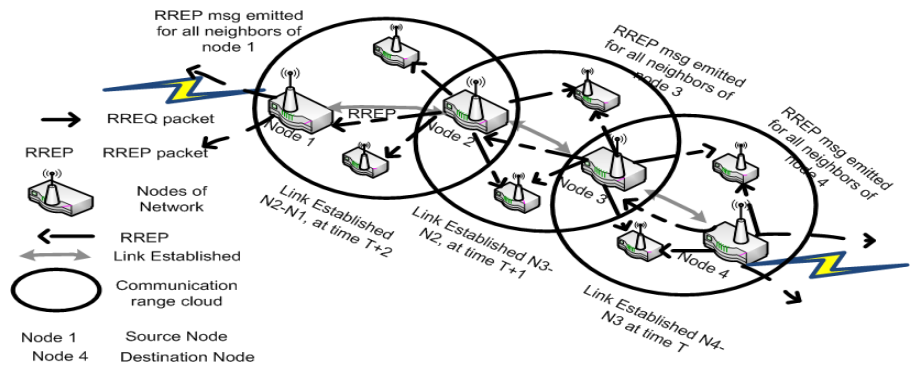
Figure 2: Route Confirmation Process(RREP)

## B. Route Maintenance

When a route is established, link is periodically monitored as shown in fig.3. If during this link sensing, routing protocol finds a broken link due to topology change or any other reason, it will generate an $RERR$ message back to main originating node of $RREQ$. When the originating node receives this $RERR$ message, it starts a new route discovery deleting the previously stored route. During route discovery time, i.e. when an $RREQ$ is broadcasted the data which actually is to be transferred from source to destination is buffered until it receives an $RREP$ packet. If $RREP$ packet is received than it is transmitted on the discovered route else, it will wait for the $RREQ_R ETRIES$ times at Maximum $TTL$. If even after that time, no $RREP$ is received than data packets are dropped [17].

Whenever a route is established via $RREQ$ and $RREP$ messages, link sensing initiates with the help of periodic messages [16] a link can be deteriorated due to noise or topology change. This is the main reason that link is being monitored periodically. In either case when a node finds no link to its next hop, it issues an $RERR$ packet informing the un-reach ability of destination node that is transmitted back to main source node. On receiving a $ERRor\ RERR$ packet, the main source node initiates new route request for broken link [35].

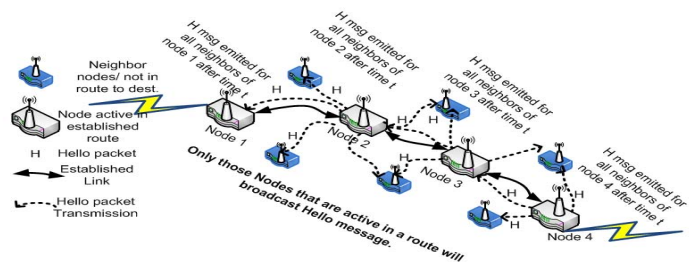
Figure 3: Link Monitoring (HELLO MESSAGE)

## C. Procedures involving Route Discovery and Maintenance

The Two main phases on which we are emphasizing in this work are route discovery and route maintenance. Discussing reactive routing, and especially $AODV$, $DSR$ and $DYMO$, Almost all these three routing protocols behave somehow in same manner. All of these use Expanding

Ring Search ($ERS$) Algorithm to avoid broadcast storm problem and Binary Exponential Back-off ($BEB$) ([15],[16],[17],[20]) Algorithm for network congestion control. Main purpose of all these routing protocols is same with differences in packet fields. Therefore it is possible to define all the said three routing protocols in a generalized algorithm which is as follows[46]:

- // initialize network
- // route required at node "n"
- Procedure route discovery RD
- Procedure $RREQ$ Forward $RREQ(f)$
- Procedure Data Packet Forward $DP(f)$
- Procedure Periodic message forward $PM(f)$
- Procedure route reply generate $RREP(g)$
- Procedure Route error message forward $RERR(f)$
- Procedure $BEB$
- Procedure $ERS$
- If $FRESH\_ENOUGH route \neq NULL$;
- Flush all $Dp(f)$
- Endif
- $FRESH\_ENOUGH route == NULL$;
- $TIMEOUT\_BUFFER \leftarrow BufferRREQ(f)$
- $TTL_S TART \leftarrow TTL$;
- Initialize Procedure $BEB$
- Initialize Procedure $ERS$
- //hop count based route discovery
- Flush all $RREQ(f)$
- for int $RREQ(f) == 1, RREQ(f) \leq TTL\_THRESHOLD, RREQ(f)^{++}$
- {
- for $TTL = 1, TTL < TTL\_THRESHOLD, TTL^{++}$
- {
- if Destination node $\leftarrow$ Receiving node
- Flush all $RREP(g)$
- if
- $RREP$ received $time < NET\_TRAVERSAL\_TIME$
- Return $RD \leftarrow Suucessful$
- break;
- initialize $RM$
- Else
- Discard $RREQ(f)$
- }
- }
- Procedure Route Maintenance $RM$
- /Link Established
- //Route Discovery Stopped
- for int $m = 1, m == number of hops, m^{++}$
- {
- for int $time = 1, time < ROUTE\_LIFE\_TIME, time^{++}$
- {

- Flush all $PM(f)$
- $PM(f) \leftarrow ROUTE\_LIFE\_TIME \div PERIODIC\_UPDATE\_TIME$
- break;
- }
- if
- $LINK_BREAKS \neq NULL$
- $RM \leftarrow$ Successful
- if $RM \geq ROUTE\_LIFE\_TIME$
- $DELETE_ROUTE$
- Elseif
- $LINKBREAKS \neq NULL$
- flush all $RERR(f)$
- initialize step $20$ to $46$
- }
- Link Established
- Flush all $DP(f)$
- End Procedures

## IV. MATHEMATICAL MODELING

Following are the major steps of Route Discovery and maintenance phases of any reactive routing protocol:

- Flooding $RREQ$ Packet (Route Request Packet)
- Receiving $RREP$ Packet (Route Reply Packet)
- Link is established and now link monitoring initiates using periodic messages
- When link is found broken, different methods apply to rectify this problem
- New route discovery/ local repair/ wait for time out occurs.

Form the above mentioned steps of Route Discovery and Route maintenance; we modeled first three steps of Reactive Routing. In this work, we have analyzed two types of scenarios i.e. the one where there is only one link active in the network and a source node $S$ wants to create a link to its destination node '$D$' during network life time '$T$'. And in other case we have tested the limits of a network of '$n$' nodes where every node is eager to send its data during network life time $T$.

Modeling route request over head, route reply overhead and hello message overhead, we follow the following scheme.

1. Network of "N" nodes Initiates
2. Route discovery $+$ Route Maintenance $==$ Routing overhead
3. Given in [21] $==$ average number of neighbors of any node in network
4. $RREQ + RREP == RouteDisco\,very overhe\,ad$
5. All number of neighbors till $i^{th}$ tier (assume dest. is at $i^{th}$) $==$ Number of $RREQ$ packets
6. $RREQ$ reaches a destination node
7. $RREP$ is generated and sent back to source node via reverse path.
8. $H ==$ number of hops from source to destination
9. Number of neighbors of all nodes including in $H$ hop $==$ number of $RREP$ packets
10. $RREP$ packet reached source node
11. Link Established

12. Route Discovery Phase Ends
13. Link maintenance phase initiates
14. Link monitoring initiates by using periodic hello messages
15. Number of active nodes/ hops in route * route life time/ periodic interval time ==Number of Hello messages. *(our enhancement)*
16. Number of $RREQ$ + number of $RREP$ + number of $HELLO$ == Routing Overhead of one route *(enhanced equation)*
17. Number of $RREQ$ for $n$ routes + number of $RREP$ for $n$ routes + number of $Hello$ packets for $n$ routes == routing over head of $n$ routes *(enhanced equation)*
18. Taking equation from point number $14$, extract parameters of $ROUTE_L IFE_T IME$ of the network and periodic hello interval.
19. Find rate of change with respect to these parameters *(our findings)*

## V. MODELING ROUTING OPERATIONS

### A. Assumptions:

- Nodes of network are placed in grid.
- Nodes have different Life Times.
- Certain sections of grid are prone to power or any other failure.
- after network initializes, there can be different variations in network parameters.

### B. Reactive Route Discovery Overhead

Route discovery overhead bears two parts i.e.
- Overhead due to $RREQ$ Propagation
- Overhead due to $RREP$ generation and propagation

Either way, control overhead of route discovery process is highly dependent upon number of hops a packet has to cross for reaching desired destination.

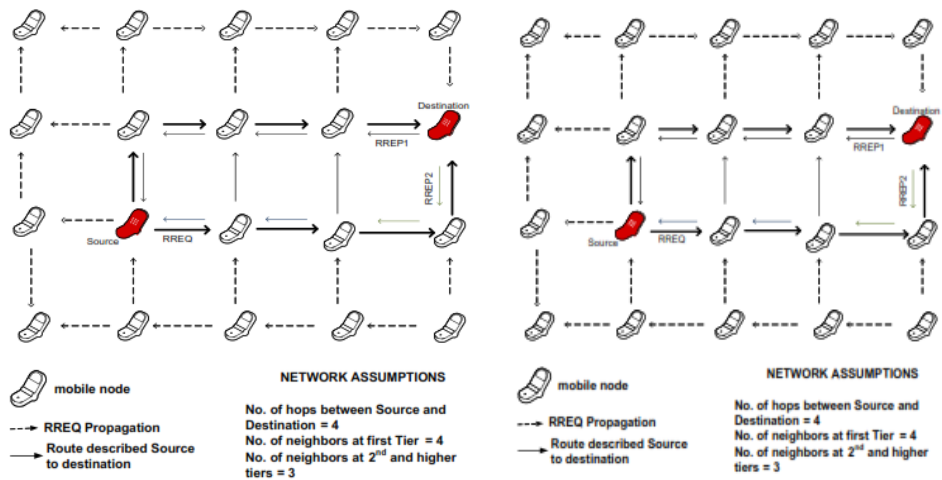

Figure 4: Propagation of $RREQ$ and $RREP$ in Network

When a node seeks a route, it propagates a route request packet in whole network until packet reaches destination node. Considering the most unfavorable scenario that source node and distinction node are placed at far corners of a network, than for sure, route request packet has to travel maximum number of hops [27]. $Fig.4$ states that when an $RREQ$ is issued from originator node, there are four neighboring nodes from second tier to $n^{th}$ tier. Moreover, a coverage index is also there at each node in between source and destination [28] to process packet. Every node that receives an $RREQ$ packet will broadcast it further ahead so that it can reach destination. If a node is receiving it for second time, packet will be discarded not broadcasted again. Authors of [21] produces mathematical framework for route discovery process combining both route request overhead and route reply overhead as:

$$R_{RREQ} = \sum_{n-1}^{H}(4)3^{H-1}\sum_{i=2}^{4}[(n-1-i)-\sum_{j=1}^{H-1}N_j]pC_i \tag{1}$$

- $C_i$ = additional coverage index of a node that has $i$ neighbors.
- $H$ = expected number of hops of network
- $N_j$ = expected number of neighbors at $^{th}$ hop.

As $RREQ$ packet finds destination node, $RREP$ is issued by destined node. This packet in the same sequence comes back to source node. To further clarify this concept, let us consider $Fig.4$ where two routes are found up to destination from source node. In the reverse fashion, two route reply packets are issued. Mathematically expected control overhead for route reply packets is given as[21]:

$$R_{RREP} = H + \frac{H}{2}(n-h-2)p \tag{2}$$

Route discovery control overhead is combination of overhead due to $RREQ$ packet and $RREP$ packet. Hence combining $Eq.1$ and $Eq.2$ we get route discovery over head.

$$R_{DISCOVERY} = RREQ + RREP \tag{3}$$

Placing values from $Eq.1$ and $Eq.2$ in $Eq.3$ we get

$$R_{DISCOVERY} = \sum_{n-1}^{H}(4)3^{H-1}\sum_{i=2}^{4}[(n-1-i)-\sum_{j=1}^{H-1}N_j]pC_i + \frac{H}{2}(n-h-2)p \tag{4}$$

### C. Reactive Route Maintenance Overhead

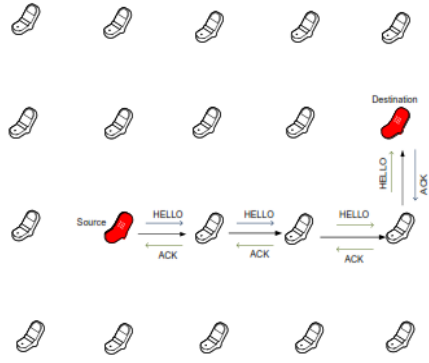

Route Monitoring Overhead
No. of Links = 4
No. of Periodic Messages per Link = 2
Route Monitoring Overhead = 4 x 2 (Route Life Time / Periodic Message Interval Time)

Figure 5: Link Monitoring Overhead of a Route

Once a route is established, link monitoring phase initiates. Purpose of link monitoring is to find out any broken link between source and distinction. If a broken link is found, that is repaired via logical link repair mechanism. For link monitoring, periodic HELLO message is propagated from all intermediate nodes that act as router for specified route. These periodic emissions of HELLO packets last until $ROUTE\_LIFE\_TIME$. In DSR, there are no $HELLO$ messages however, $ACK$ messages works almost in same manner ([16],[17]). As it is understood that link life time of any route is a random variable and can have a life of between route establishment and route expiry. Mostly, a link is prone to breaks in crucial environments (high mobility or high scalability or both)[29].

$Fig.5$ that represents an established route between two nodes and link monitoring messages illustrates that when a route that have only one link will have only one periodic $HELLO$ message for link monitoring purpose till route is expired. $Eq.5$ depicts the routing load of link monitoring messages for one route.

$$R_{HELLO(e)} = 2(\frac{T}{t})l \tag{5}$$

- $R_{HELLO(e)}$ = Number of HELLO messages for monitoring single route
- $T$ = Route Life Time, the time after that route is expired
- $t$ = Periodic interval time of HELLO messages

$Eq.5$ expresses the routing load considering link monitoring of a single route. To calculate routing load of $n$ routes of a network, we get

$$R_{HELLO} = \sum_{i=1}^{n} 2(\frac{T}{t})l \tag{6}$$

### D. Aggregate Reactive Overhead

Aggregate routing overhead can be termed as the sum of routing overhead due to route discovery and overhead due to link monitoring of a route as shown in $Eq.7$

$$RO = R_{DISCOVERY} + R_{HELLO} \tag{7}$$

Placing values from $Eq.4$ and $Eq.6$, we get:

$$RO = \sum_{n-1}^{H}(4)3^{H-1}\sum_{i=2}^{4}[(n-1-i)-\sum_{j=1}^{H-1}N_j]p(_i)+H+\frac{H}{2}(n-h-2)p+\sum_{i=1}^{n}2(\frac{T}{t})l \quad (8)$$

$$RO = \sum_{n-1}^{H}(4)3^{H-1}\sum_{i=2}^{4}[(n-1-i)-\sum_{j=1}^{H-1}N_j]p(_i)+H+\frac{H}{2}(n-h-2)p+\sum_{i=1}^{n}2(\frac{T}{t})l \quad (9)$$

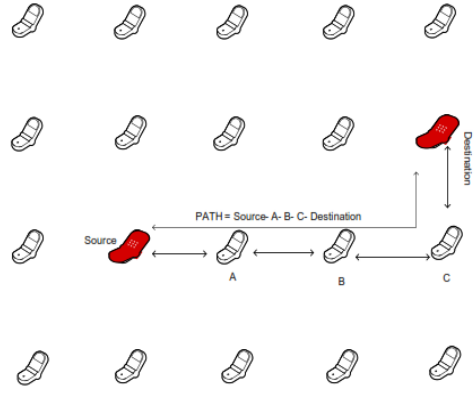

Figure 6: Source to Destination Route(Grid Environment)

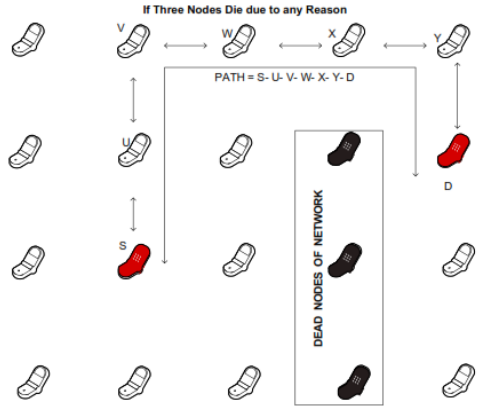

Figure 7: Source to Destination Route(Portion of Grid Black Out)

## VI. MODELING ROUTING VARIATIONS

In our proposed model, as network nodes are placed in a grid where different sections of grid are vulnerable to failure. Such failures results in unpredictability of network. The nodes that initially are placed in a grid are then mobile after network initialization. Adding further, these nodes have different life times as well. Hence our model for reactive routing is totally unpredictable and variable.

Focusing *Fig*.3 and *Eq*.8 that reflects control overhead due to route discovery and monitoring phases of a reactive routing protocol, it is obvious that if there is any variation in node density a new route is to be established. That new route may have different number of hops and intermediate nodes. moreover, there are different node life times, hence to impact of all these variations on overall control overhead, we need to find rate of change within these parameters with respect to *Eq*.8.

To calculate overhead in such variable network whose almost all parameters may change at any instance we get our function $y$ from *Eq*.8 and to find rate of change in different network parameters, we undertook calculations by using partial derivations. We consider number of nodes, number of hops, route life time and periodic message interval time as network parameters that can vary. In further sections we give analytical model reflecting variability of said parameters in a network.

$$y(n,H,T,t) = \sum_{n-1}^{H}(4)3^{H-1}\sum_{i=2}^{4}[(n-1-i)-\sum_{j=1}^{H-1}N_j]p.C_i) + H + \frac{H}{2}(n-h-2)p \quad (10)$$

Where
- $n$ = number of nodes in a network
- $H$ = number of hops of a network
- $T$ = route life time while
- $t$ = periodic interval time for link monitoring.

*Eq*.9 reflects that number of links in a route can be dependent on number of nodes of network however, number of link monitoring messages i.e. HELLO messages are dependant upon number of hops of a route, route life time and periodic HELLO message interval time.

### A. Variation in Scalability

We take partial derivative with respect to $n$ to calculate variation in number of nodes in a network and we get our *Eq*.10:

$$\partial y/\partial n = \sum_{n-1}^{H}(4)3^{H-1}[\sum_{i=2}^{4}[(-i)-\sum_{j=1}^{H-1}N_j]p.C_i] + H + \frac{H}{2}(-h-2)p \quad (11)$$

Variation in number of nodes considerably effects number of hops for certain routes within network. Likewise, changing number of hops of a route surely effects link monitoring overhead. To analyze rate of change with respect to hops we get:

$$\partial y/\partial H = \sum_{n-1}^{H}(4)3^{H-1} + H - 1(3^{H-1})[\sum_{i=2}^{4}[(n-1-i)-\sum_{j=1}^{H-1}N_j]p.C_i] + 1 + \frac{1}{2}(n-3)p \quad (12)$$

As discussed earlier, number of nodes and number of hops play vital role in control overhead. to find overall control overhead we use chain rule assuming rest of chosen parameters as constant. Taking *Eq*.4 as a function $x$ we calculate overall rate of change with respect to $n$ and $H$ as *Eq*.10 and *Eq*.11 respectively.

$$dx = (\frac{\partial x}{\partial n})dn + (\frac{\partial x}{\partial H})dH \quad (13)$$

Substituting values in Eq.12:

$$dx = (\sum_{n-1}^{H}(4)3^{H-1}[\sum_{i=2}^{4}[(n-1-i)-\sum_{j=1}^{H-1}N_j]p(C_i)] + H + \frac{H}{2}(n-h-2)p)dn +$$

$$(\sum_{n-1}^{H}(4)3^{H-1}[\sum_{i=2}^{4}[(n-1-i)-\sum_{j=1}^{H-1}N_j]p(C_i)]+1+\frac{1}{2}(n-3)p)dH$$

## B. Variation in Route Life Time and Link Monitoring

Calculating rate of change in route life time with respect to function $y$:

$$\partial y/\partial T = \sum_{i=1}^{n}(\frac{2}{t})l_i \quad (14)$$

And to compute variation in periodic interval time for link monitoring:

$$\partial y/\partial t = \sum_{i=1}^{n}-2(\frac{T}{t^2})l_i \quad (15)$$

$Eq.14$ and $Eq.15$, states that if there are different active routes with different active periodic message intervals in a network, than, overall control overhead is total derivative w.r.t. route life time and periodic message interval time. To analyze this, we consider $R_{HELLO}$ (expressed in $Eq.6$) as a function $z$ whose partial derivatives are expressed in $Eq.14$ and $Eq.15$. by applying total derivation, we find our $Eq.16$

$$dz = (\frac{\partial z}{\partial T})dT + (\frac{\partial z}{\partial t})dt \quad (16)$$

Now substituting values to get control overhead resulting from varying parameters of route life time and periodic message update time:

$$dz = (\sum_{i=1}^{n}(\frac{2}{t})l_i)dT + (\sum_{i=1}^{n}-2(\frac{T}{t^2})l_i)dt \quad (17)$$

## C. Variation in Over all Network Parameters

To give an optimum model for control overhead of route discovery and route monitoring respecting reactive routing, we apply chain rule on function $y$ to get sum of all partial derivatives of a function in $Eq.19$

$$dy = (\frac{\partial y}{\partial n})dn + (\frac{\partial y}{\partial H})dH + (\frac{\partial y}{\partial T})dT + (\frac{\partial y}{\partial t})dt \quad (18)$$

Placing the values, we get:

$$dy = (\sum_{n-1}^{H}(4)3^{H-1}[\sum_{i=2}^{4}[(n-1-i)-\sum_{j=1}^{H-1}N_j]p(C_i)]+H+\frac{H}{2}(n-h-2)p)dn +$$
$$(\sum_{n-1}^{H}(4)3^{H-1}[\sum_{i=2}^{4}[(n-1-i)-\sum_{j=1}^{H-1}N_j]p(C_i)]+1+\frac{1}{2}(n-3)p)dH +$$
$$(\sum_{i=1}^{n}(\frac{2}{t})l_i)dT + (\sum_{i=1}^{n}-2(\frac{T}{t^2})l_i)dt \quad (19)$$

# VII. SIMULATION RESULTS AND DISCUSSIONS

We use $NS-2$ as our simulation tool. $AODV$ [31] coding was developed by CMU/MONARCH group while it was optimized by Samir Das and Mahesh Marina (University of Cincinnati). Coding of $DYMOUM$ by $MASIMUM$ [33] is used for $DYMO$. We use $NS-2.34$ for simulating $AODV$ and $DSR$ while, $DYMOUM$ is simulated in $NS-2.29$. We focus on the mobility and scalability factors of Ad Hoc networks in our work.

We considered a network of 50 nodes where nodes are randomly located and are mobile. These nodes have a bandwidth of $2Mbps$ each. Mobility is set as $2m/s$ which is average walking speed. Packet size is defined as $512 bytes$, while simulation setup runs on Continues Bit Rates ($CBR$). The size of network is defined as $1000m^2$. Given these parameters, we have confined our experiments to following three metrics.

1. Throughput
2. Delay
3. Routing Load

## A. Throughput of Reactive Protocols

In general sense, throughput refers to the amount of data that has successfully reached its destination. Mathematically it can be stated as:

$$Througput = \frac{messagesRecievedSuccessfully}{Time} \qquad (20)$$

*Mobility Factor:* Considering graph ($fig.8$) for throughput, $DSR$ attains the maximum throughput with respect to $AODV$ and $DYMO$. If we consider $AODV$, than it, surely have a $TIMEOUT$ factor involved. $AODV$ waits for a specified time, then route is termed invalid and finally erased from routing table. ``$HELLO$'' messages (used for link monitoring) in $AODV$ also works very well for mobile environment. Overall considering mobility factor, $DSR$ gives stable throughput, as no unnecessary packets are generated by this routing protocol. In link breakages, $DSR$ have multiple routes while, in $AODV$, routing table keeps the best chosen path only. Hence, within the environment where links are immune to breaks, $DSR$ supersedes $AODV$ and $DYMO$. $DYMO$ proves to be the worst amongst the other two protocols.

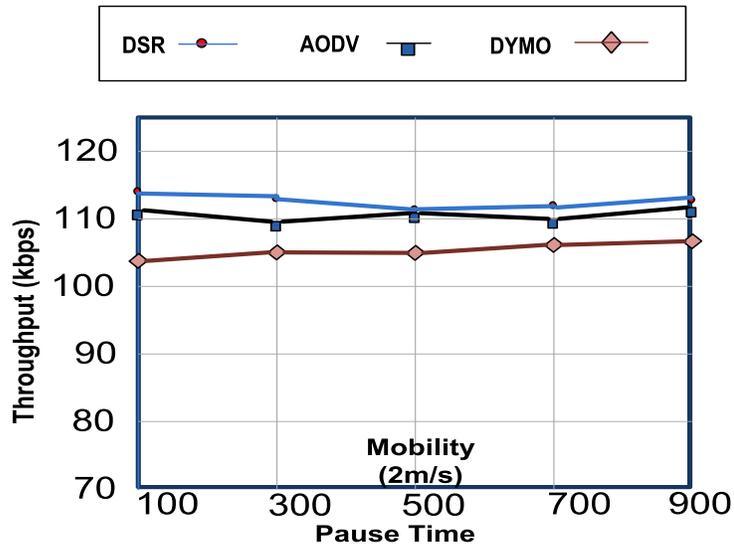

Figure 8: Throughput(Mobility): $AODV, DSR, DYMO$

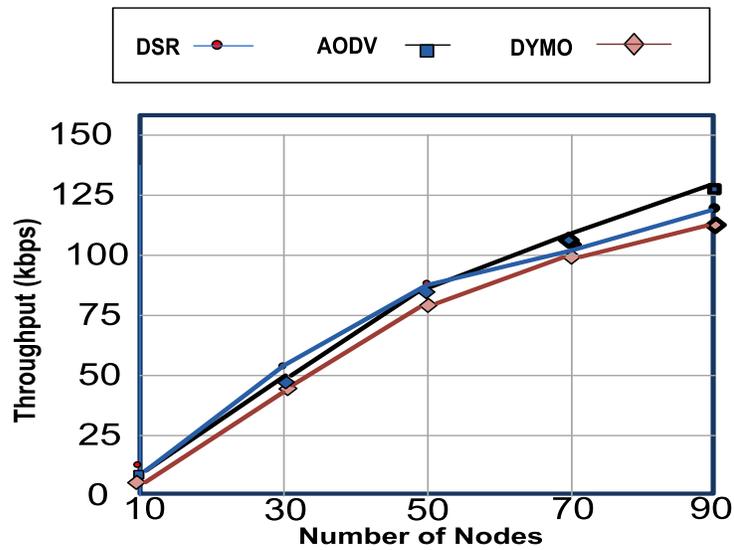

Figure 9: Throughput(Scalability): $AODV, DSR, DYMO$

*Scalability Factor:* According to experiments performed $(fig.9)$, $AODV$ converges at almost all data rates with salabilities. While $DSR$ proves itself to be scalable but only during high data traffic, it can not converge the network. $DYMO$ performs worst among these studied routing protocols. As the number of nodes increases or data traffic increases, its performance degrades dramatically. According to [30], a network of multiple thousands of nodes with different traffic loads can be handled by $AODV$. The reason that $AODV$ supersedes $DSR$ and $DYMO$ is lower packet loss ratio and propagation of information regarding distant vector which practically consume minimum bandwidth. This feature gives $AODV$ a room for scalability. In $AODV$, routing packet contains only one hop information while in $DSR$, packet size is larger as it keeps the information of whole route. This is another reason that $AODV$ outperforms $DSR$.

## B. End to End Delay of Reactive Routing

Time which a packet takes in reaching destined node from the originator node can be termed as end to end delay. Mathematically we can express it as:

$$ED = \frac{(Number of Transmitted Packets)(RTT)}{Number of Recieved Packets}$$

*Mobility Factor:* As shown in the graph $(fig.10)$, $AODV$ gives lowest performance as, link breakages may lead to longer routes. $DYMO$, though works worst in throughput case but here it works best amongst $DSR$ and $AODV$. It is so because, $DYMO$ does not check the routes in memory as $DSR$ looks into route cache and $AODV$ in to its routing table, instead it starts Expanding Ring Search $(ERS)$ algorithm whenever a route is required.

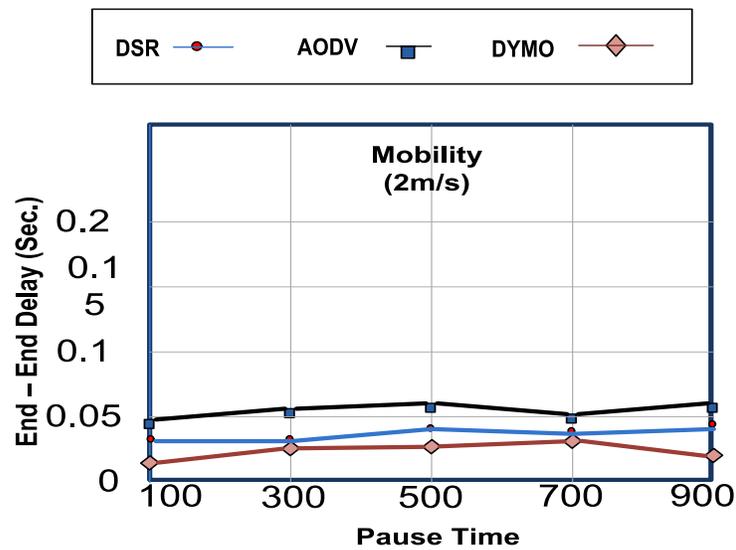

Figure 10: End to End Delay (Mobility): $AODV, DSR, DYMO$

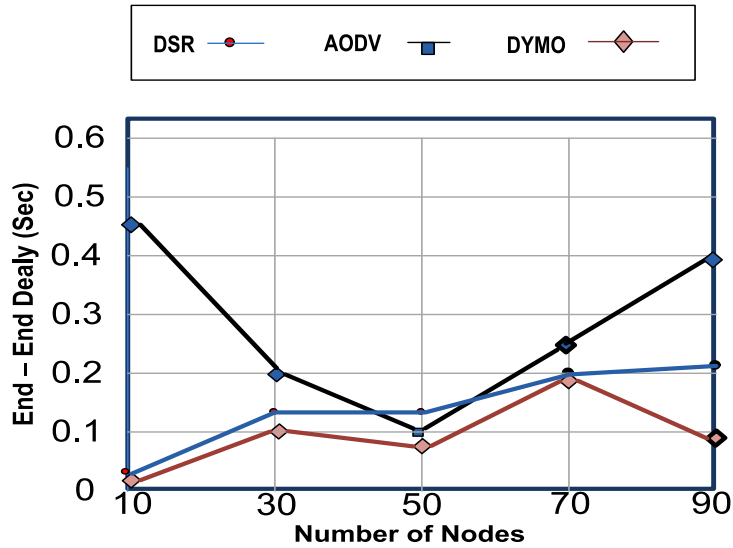

Figure 11: End to End Delay (Scalability): $AODV, DSR, DYMO$

*Scalability Factor:* Keenly observing $fig.11$ we can infer that due to the concept of *Grat. RREP* in *DSR, AODV* and *DYMO* results in lowest End to End delay, irrelevant of number of nodes in the network. $Gratitous RREP$ though results in lower delay at normal traffic rates though, $DSR$ checks the route cache before starting Expanding Ring Search $(ERS)$ algorithm in the same way as $AODV$ search route in its routing table before starting a route request using $ERS$ Algorithm. $DYMO$ does not use such stored information rather it simply initiates $ERS$. $AODV$ also have a link repair feature that makes it bear the highest end to end delay with respect to any scalability among $DYMO and DSR$.

### C. Routing Load of Reactive Routing

When a single data packet is to be sent from one node to another within a network, a number of routing packets are involved in sending this data packet. The numbers of these routing packets which are sent just to transfer one data packet are termed as Routing Load or Normalized Routing Load. Mathematically, we can state:

$$RoutingLoa\ d = (Routing\ + DataLoad\ ) - (NumberofDa\ taPacketss\ ent)$$

*Mobility Factor:* $AODV$ and $DSR$ use the concept of $grat.RREP$, i.e. when a $RREQ$ reaches any node that has a valid route stored in its route cache or routing table, it generates a $RREP$ by itself to the original source node. This $RREP$ contains the full information up to the destined node and overhead of finding route beyond that node limits. $DYMO$ does not use this $grat.RREP$. That's why it suffers from greater routing overhead with respect to the other two protocols. $AODV$ also works well in the context of normalized routing overhead however, there is a concept of local link repair and above all, use of $HELLO$ message for link monitoring, makes it performance lower then $DSR$. A node with underlying $DSR$ protocol use promiscuous mode and this is the reason that it bears lowest overhead $(fig.12)$.

A common observation with respect to increase in mobility of nodes in the network is that all the three routing protocols bear gradually higher overhead. The reason is propagation of route error packets. As the mobility increases, chances of link breaks also increase in the same proportion which results in increase of routing overhead.

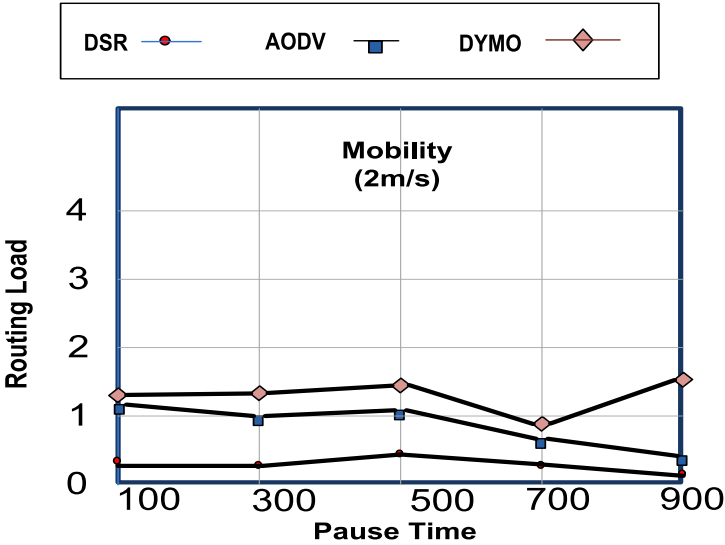
Figure 12: Routing Load (Mobility): AODV, DSR, DYMO

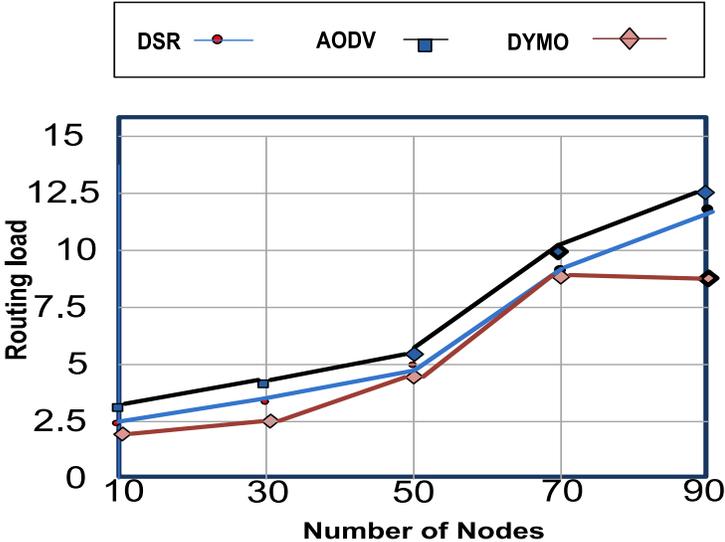
Figure 13: Routing Load (Scalability): AODV, DSR, DYMO

*Scalability Factor:* Routing overhead of $DYMO$ is lower than that of $AODV$ and $DSR$ ($fig.13$). $AODV$ bears high routing overhead in dense networks. Periodic link sensing packets involved in local link repair mechanism and $grat.RREP$ results in high routing overhead. Whereas promiscuous mode utilized by $DSR$ reduces the routing overhead in not so dense environment.

## VIII. PERFORMANCES AND COMPARISONS

The protocol that uses minimum resources of bandwidth by its control packets can provide better data flow. Hence, the environments where traffic load is very high, protocols having low routing overhead survive. If we consider scalability, than $AODV$ stands at top of rest of studied routing protocols. It uses distance vector distribution that minimize network resource consumption. The network underlying $AODV$ protocol bears low routing overhead as control packets of $AODV$ contains a very small part of information in them where as if we compare it with $DSR$, control packet of $DSR$ carries whole routing information in it. Hence we can say that $DSR$ has higher routing overhead in terms of bytes or size. If we consider number of control packets than $DSR$ broadcast less number of packets than that of $AODV$. $AODV$ use periodic hello packet for link sensing and also bear local repair routing overhead. Hence if we compare both of these routing protocols ($AODV$ and $DSR$) considering mobility and speed factors, we can conclude that both of these protocols give more or less same performance.

Concluding all the routing protocols, our study suggest that, $AODV$ can be selected for denser environments where lower routing overhead is required, $DSR$ should be used within a network having limited number of hops but it is better for highly mobile environment. $DYMO$ routing protocol can be used in networks where delay is in tolerable. As like other reactive protocols, $DYMO$ does not look for any stored route as $DSR$ looks into its cache and $AODV$ in its routing table. It initializes binary exponential back off and $ERS$ algorithm immediately.

Observing simulated results keenly, we can deduce that, $AODV$ enjoys higher throughput on the cost of longer delay and increased routing load. To maintain link connectivity, every node in a route propagates periodic HELLO message. In case of broken link, logical link repair initiates that obviously don't allow packet to be dropped but results in increased routing overhead plus longer delay.

If we consider a network underlying $DSR$ routing protocol, $DSR$ enjoys higher throughput by paying price of longer end to end delay. When a route is required, $DSR$ initially finds it in route cache. If no route is found than, route request packet is propagated for required destination. This process leads to end to end delay however, ensures throughput. Considering throughput, $DYMO$ performs worst as it gives least delay however, this reduces delay time is compromised with packet drop ratio. The less delay $DYMO$ enjoys more packet drop ratio it bears.

### A. Tabular Representation

Given tables very clearly explain the findings of our simulated results. In these tables we give a brief comparison analysis of studied three routing protocols i.e. $AODV$, $DSR$ and $DYMO$. This comparison is solely based upon simulated results for the said reactive routing protocols.

Table. 1 gives general differences and techniques being used in these three most prominent reactive routing protocols. In Table 2, a comparison is made amongst $AODV$, $DSR$ and $DYMO$ considering mobility factor. $AODV$ and $DSR$ has better throughput however, they have to compromise on end to end delay time for this higher throughput. However, $DYMO$ though has a bit less throughput however, there is no delay.

Table 3 discusses the different mobility sceneries and categorize these three routing protocols as best, average and worst with respect to throughput, delay and routing load [32].

Scalability in $AODV, DSR and DYMO$ is discussed in Table 4 that states that, $AODV$ stands best considering throughput metric amongst $DSR$ and $DYMO$.

Table 1: Basic Features: Reactive Routing Protocols

| Feature | AODV | DSR | DYMO |
|---|---|---|---|
| Protocol type | Distance Vector | Source routing | Source routing |
| Route maintained in | Routing table | Route Cache | Routing table |
| Multiple route discovery | No | Yes | No |
| Update destination | Source | Source | Source |

| Broadcast | Full | Full | Full |
|---|---|---|---|
| Reuse of routing information | No | Yes | No |
| Route selection | Only searched route | Hop count | Only searched route |
| Route reconfiguration | Erase route notify source | Erase route notify source | Erase route notify source |
| Route discovery packets | using RREQ and RREP packets | using RREQ and RREP packets | using RREQ and RREP packets |
| Limiting overhead, collision avoidance, network congestion | Expanding Ring Search Algorithm | Expanding Ring Search Algorithm | Expanding Ring Search Algorithm |
| Limiting overhead, collision avoidance, network congestion | Binary Exponential Back off Time | Binary Exponential Back off Time | Binary Exponential Back off Time |
| Update information | By RERR message | By RERR message | By RERR message |

Table 2: Comparison Reactive Protocols w.r.t. Mobility

| Protocol | Routing Tech. | Pro's | Con's |
|---|---|---|---|
| AODV | Seq. Number with Logical Link Repair | Better Throughput | Delay due to LLR |
| DSR | Route Cache-ing | Memorizing Routes and Better Throughput | Delay at high mobility |
| DYMO | Without Route Cache and *Grat.* RREP | Minimize Delay in high Mobility | Low Throughput at high mobility |

Table 3: Performance of Reactive Protocols at different Speeds

| Mobility | Protocol Performing | Delay | Routing Load | Through put |
|---|---|---|---|---|
| High Mobility (0-300s) Pause Timings | Best ⇒ | ADOV | DYMO | DSR |
| | Average ⇒ | DSR | AODV | AODV |
| | Worst ⇒ | DYMO | DSR | DYMO |
| Avg. Mobility (300-700s) Pause Timings | Best ⇒ | DSR | DYMO | DSR |
| | Average ⇒ | AODV | AODV | AODV |
| | Worst ⇒ | DYMO | DSR | DYMO |
| Low Mobility (700-900s) Pause Timings | Best ⇒ | DYMO | DYMO | AODV |
| | Average ⇒ | DSR | AODV | DSR |
| | Worst ⇒ | AODV | DSR | DYMO |
| Mixed Mobility (0-900s) Pause Timings | Best ⇒ | AODV | DYMO | DSR |
| | Average ⇒ | DSR | AODV | AODV |
| | Worst ⇒ | DYMO | DSR | DYMO |

Table 4: Comparison Reactive Protocols w.r.t. Scalability

| Protocol | Routing Tech. | Pro's | Con's |
|---|---|---|---|
| AODV | Periodic Link Monitoring, grat. RREP | Best Throughput | Maximum Delay due to LLR |
| DSR | Route Cache-ing | Lower Delay | higher Routing Load |
| DYMO | Without Route Cache and *Grat.* RREP | Low Delay | High Routing Load |

## II. CONCLUSION

This work is the enhancement of our previous work where we present control overhead frame work for route request, route reply and link monitoring processes. After calculating aggregate routing overhead, we took different metrics as number of nodes, number of hops per route, route life time, and periodic interval of link monitoring messages and occurring frequency of trigger messages. These parameters are varied to express the impact of their variation in network. In next phase of our work, we simulated $AODV$, $DSR$ and $DYMO$ for their performance analysis and comparisons with respect to mobility and scalability concerns. These experiments are discussed in graphical, textual and tabular forms to present a better picture and understanding of functionality of these three reactive protocols. We confine our selves to mobility and scalability aspects keeping metrics of throughput, delay and control over head. Our simulated results show that network running over $AODV$ has lower overhead with respect to $DSR$ based network as $DSR$ has to carry extra bytes of source routes as well. Contrary to this, $DSR$ has lower routing overhead if we consider only number of packets. $AODV$ and $DSR$ performs best in all mobility and scalability scenarios however, if we have some non delay tolerant network, $DYMO$ is the protocol that must be used.